\documentclass[%
    10pt,
    prx,
    superscriptaddress,
    a4paper,
    twocolumn,
    nofootinbib,
    nobibnotes,
    floatfix,
    longbibliography,
    tightenlines,
    showkeys
]{revtex4-2}

\usepackage[american]{babel}
\usepackage[utf8x]{inputenc}

\usepackage{grffile} 

\usepackage[scaled=.93]{newtxtext}
\usepackage[bigdelims, scaled=1.0,vvarbb,subscriptcorrection]{newtxmath}
\usepackage[scr=boondoxo,cal=boondoxo]{mathalfa}

\usepackage[scaled=.85]{FiraSans}

\usepackage{bbm}
\usepackage{mathtools}
\usepackage{dsfont}
\usepackage{braket}
\usepackage{cancel} 

\usepackage{graphicx}
\usepackage{dashrule}

\usepackage{siunitx} 
\usepackage{xfrac}
\usepackage{placeins}
\usepackage[clock]{ifsym}
\usepackage{esvect}

\usepackage{ragged2e}
\usepackage{array}

\usepackage{booktabs}
\usepackage{tabularx,colortbl}
\usepackage{multirow}
\usepackage[table]{xcolor}
\usepackage{adjustbox}

\renewcommand{\vec}[1]{\mathbf{#1}}

\newcommand{\ii}{\mathrm{i}}

\newcommand{\dd}{\!\mathrm{d}}

\usepackage{graphicx}

\usepackage{xcolor}
\definecolor{cset-aps-blueberry}{RGB}{28,128,158}
\definecolor{cset-aps-blue}{RGB}{46,44,184}
\definecolor{cset-aps-turquoise}{RGB}{0,67,88}
\definecolor{cset-aps-limegreen}{RGB}{190,219,67}
\definecolor{cset-aps-green}{RGB}{31,138,112}
\definecolor{cset-aps-yellow}{RGB}{255,225,25}
\definecolor{cset-aps-orange}{RGB}{253,116,0}
\definecolor{cset-aps-red}{RGB}{219,0,43}
\definecolor{cset-aps-violett}{RGB}{142,68,173}

\definecolor{cset-aps-kobalt-medium}{RGB}{62,54,222}
\definecolor{cset-aps-kobalt-dark}{RGB}{28,24,150}

\definecolor{cset-aps-my-label-red}{RGB}{202,0,17}
\definecolor{cset-aps-my-label-blue}{RGB}{53,71,140}
\definecolor{cset-aps-my-label-gray}{RGB}{145,145,145}

\usepackage{hyperref}
\hypersetup{%
    colorlinks=true,
    linkcolor={cset-aps-red},
    linkbordercolor={cset-aps-red},
    filecolor={cset-aps-orange},
    filebordercolor={cset-aps-orange},
    citecolor={cset-aps-kobalt-medium},
    citebordercolor={cset-aps-kobalt-medium},
    urlcolor={cset-aps-kobalt-dark},
    urlbordercolor={cset-aps-kobalt-dark},
    menucolor={cset-aps-limegreen},
    menubordercolor={cset-aps-limegreen},
    breaklinks=true,
    pdfborderstyle={/S/U/W 2},
    pdfpagemode=UseOutlines,
    pdfstartpage={1},
}

\usepackage{braket}

\patchcmd{\section}
  {\centering}
  {\sffamily\firabook\bfseries\normalsize\raggedright}
  {}
  {}

\patchcmd{\subsection}
  {\centering}
  {\sffamily\firabook\bfseries\normalsize\raggedright}
  {}
  {}

\makeatletter
\renewcommand*{\fnum@figure}{{\normalfont\sffamily {\bfseries\MakeUppercase{\figurename~\thefigure}}}}
\renewcommand*{\@caption@fignum@sep}{{\sffamily:} }
\makeatother

\newcommand{\iii}{{(\alpha)}}
\renewcommand{\i}{{(1)}}
\renewcommand{\ii}{{(2)}}
\newcommand{\pathlabel}{\alpha}
\newcommand{\pathindex}{(\pathlabel)}

\newcommand{\statelabelExcited}{a}
\newcommand{\statelabelGround}{b}

\newcommand{\keff}{~k_m}

\newcommand{\affHAN}{\address{Institut f{\"u}r Quantenoptik, Leibniz Universit{\"a}t Hannover, \\
    Welfengarten 1, D-30167 Hannover, Germany}}
\newcommand{\affULM}{\address{Institut f{\"u}r Quantenphysik and Center for Integrated Quantum
    Science and Technology (IQ\textsuperscript{ST}), Universit{\"a}t Ulm, Albert-Einstein-Allee 11, D-89069 Ulm, Germany}}
\newcommand{\affZYCH}{\address{Centre for Engineered Quantum Systems, School of Mathematics and Physics, The University of Queensland, St Lucia, QLD 4072, Australia}}
\newcommand{\affSCHLEICH}{\address{Hagler Institute for Advanced Study and Department of Physics and Astronomy, Institute for Quantum Science and Engineering (IQSE), Texas A{\&}M AgriLife Research, Texas A{\&}M University, College Station, TX 77843-4242, USA}}
\newcommand{\affDLR}{\address{Institute of Quantum Technologies, German Aerospace Center (DLR), Wilhelm-Runge-Straße 10, D-89081 Ulm, Germany}}

\begin{document}
\title[Interference of Clocks]{%
{\sffamily\normalfont\firalight\LARGE\scshape Interference of Clocks: A Quantum Twin Paradox}
\\[1ex]
\small\normalfont{Published in \href{https://doi.org/10.1126/sciadv.aax8966}{Science Advances, Vol. 5, No. 10, eaxx8966 (2019)}}%
}

\author{Sina Loriani $^\dagger$}
\affHAN
\author{Alexander Friedrich $^\dagger$}
\affULM
\email{alexander.friedrich@uni-ulm.de}
\author{Christian Ufrecht} 
\affULM
\author{Fabio Di Pumpo}   
\affULM
\author{Stephan Kleinert} 
\affULM

\author{Sven Abend}
\affHAN
\author{Naceur Gaaloul}
\affHAN
\author{Christian Meiners}
\affHAN
\author{Christian Schubert}
\affHAN
\author{Dorothee Tell}
\affHAN
\author{\'Etienne Wodey}
\affHAN

\author{Magdalena Zych}
\affZYCH

\author{Wolfgang Ertmer}
\affHAN
\author{Albert Roura} 
\affDLR
\author{Dennis Schlippert}
\affHAN
\author{Wolfgang P. Schleich}
\affULM 
\affSCHLEICH
\author{Ernst~M.~Rasel} 
\affHAN
\author{Enno Giese}
\affULM

\begin{abstract}
\begin{center}
$^\dagger$ These authors have contributed equally to this work.\\[1ex]
\end{center}
\noindent The phase of matter waves depends on proper time and is therefore susceptible to special-relativistic (kinematic) and gravitational (redshift) time dilation.
Hence, it is conceivable that atom interferometers measure general-relativistic time-dilation effects.
In contrast to this intuition, we show: (i.) Closed light-pulse interferometers without clock transitions during the pulse sequence are not sensitive to gravitational time dilation in a linear potential.
(ii.) They  can constitute a quantum version of the special-relativistic twin paradox. 
(iii.) Our proposed experimental geometry for a quantum-clock interferometer isolates this effect.
\end{abstract}

\maketitle

\section*{Introduction}

\noindent Proper time is operationally defined \cite{Einstein05} 
as the quantity measured by an ideal clock~\cite{Moeller56} moving through spacetime.
As the passage of time itself is relative, the comparison of two clocks that travelled along different worldlines gives rise to the twin paradox~\cite{Einstein18}.
Whereas this key feature of relativity relies on clocks localised on worldlines, today's clocks are based on atoms that can be in a superposition of different trajectories.
This nature of quantum objects is exploited by matter-wave interferometers which create superpositions at macroscopic spatial separations~\cite{Kovachy15}.
One can therefore envision a \emph{single} quantum clock such as a two-level atom in a \emph{superposition} of two different worldlines, suggesting a twin paradox, in principle susceptible to any form of time dilation~\cite{Sinha11,Zych11,Zych17}.
We demonstrate which atom interferometers implement a quantum twin paradox, how quantum clocks interfere, and their sensitivity to different types of time dilation.

The astonishing consequences of time dilation can be illustrated by the story of two twins~\cite{Einstein18}, depicted in Fig.~1A:
Initially at the same position, one of them decides to go on a journey through space and leaves his brother behind.

\begin{figure}
\centering
\begin{center}
\includegraphics[width=\columnwidth]{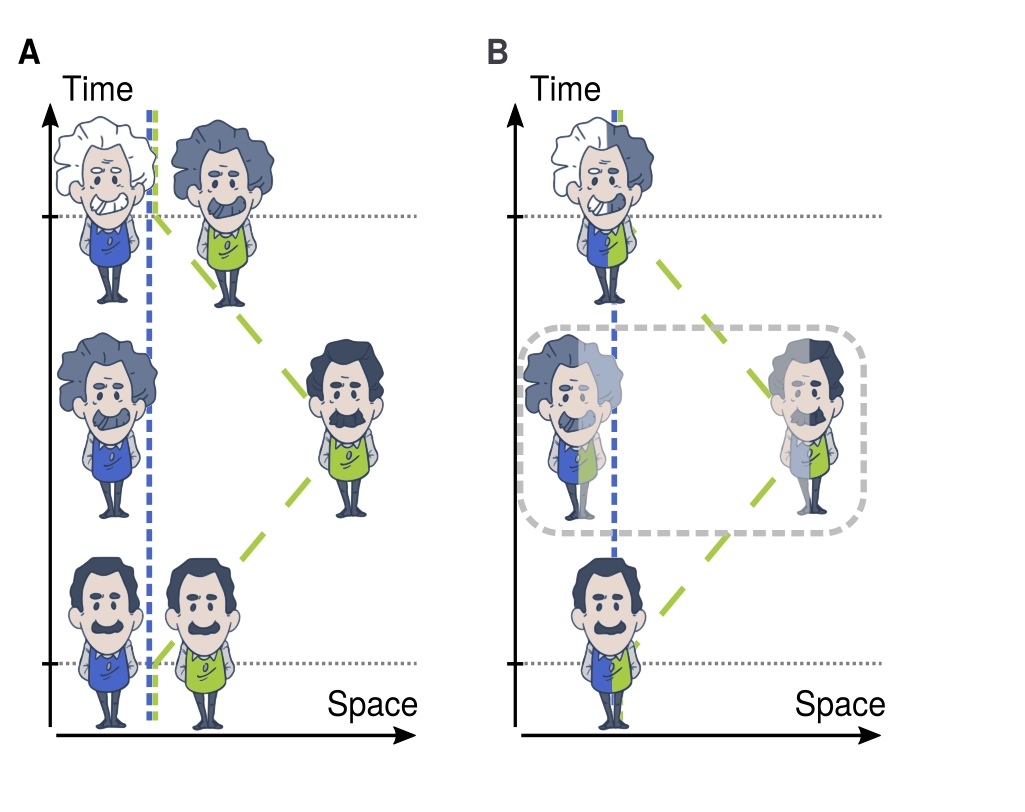}
\end{center}
\caption{{\sffamily\bfseries\firamedium Twin paradox and its quantum version.} 
(\textbf{{\sffamily A}})~As a consequence of relativity, two initially co-located twins experience time dilation when traveling along different worldlines. Upon reunion, they find that they aged differently due to the relative motion between them.
(\textbf{{\sffamily B}})~In a quantum version of this gedankenexperiment, a single individual is travelling along two paths in superposition, serving as its own twin and aging at two different rates simultaneously.}
\label{fig:twins}
\end{figure}

Due to their relative motion, he experiences time dilation and, upon meeting his twin again after the voyage, has aged slower than his brother who remained at the same position. Even though this difference in age is striking by itself, the twin who travelled could argue that from his perspective, his brother has moved away and returned, making the same argument. This twin paradox can be resolved in the context of relativity, where it becomes apparent that not both twins are in an inertial system for the whole duration. In the presence of gravity two twins which separate and reunite experience additional time dilation depending on the gravitational potential during their travel.

The experimental verifications of the effect that leads to the difference in age, namely special-relativistic and gravitational time dilation, were milestones in the development of modern physics and have, for instance, been performed by the comparison of two atomic clocks~\cite{Hafele72b,Vessot80,Chou10}.
Atomic clocks, as used in these experiments, are based on microwave and optical transitions between electronic states and define the state of the art in time keeping~\cite{Nicholson15}.

In analogy to optical interferometry, atom interferometers measure the relative phase of a matter wave accumulated during the propagation by interfering different modes.
Even though it is possible to generate these interferometers through different techniques, we focus here on light-pulse atom interferometers like the one of Kasevich and Chu~\cite{Kasevich91} with two distinct spatially separated branches, where the matter waves are manipulated through absorption and emission of photons that induce a recoil to the atom.
Conventionally, such interferometers consist of a series of light pulses that coherently drive atoms into a superposition of motional states, leading to the spatial separation.
The branches are then redirected and finally recombined, such that the probability to find atoms in a specific momentum state displays an interference pattern and depends on the phase difference $\Delta \varphi$ accumulated between the branches that is susceptible to inertial forces. 
As such, light-pulse atom interferometers do not only provide high-precision inertial sensors~\cite{Freier16,Savoie18} with applications in tests of the foundations of physics~\cite{Bouchendira11,Schlippert14,Zhou15,Barrett16,Rosi14,Overstreet18,Parker18}, but at the same time constitute a powerful technique to manipulate atoms and generate spatial superpositions.

Atom interferometry in conjunction with atomic clocks has led to the idea of using time dilation between two branches of an atom interferometer as a which-way marker to measure effects like the gravitational redshift through the visibility of the interference signal~\cite{Sinha11,Zych11}.
However, no specific geometry for an atom interferometer was proposed and no physical process for the manipulation of the matter waves was discussed.
The geometry as well as the protocols used for coherent manipulation crucially determine if and how the interferometer phase depends on proper time~\cite{Giese19}.
Therefore, the question whether effects connected to time dilation can be observed in light-pulse atom interferometers is still missing a conclusive answer.

In this work, we study a quantum version of the twin paradox, where a single twin is in a superposition of two different worldlines, aging simultaneously at different rates, illustrated in Fig.~1B.
We show that light-pulse atom interferometers can implement the scenario where time dilation is due to special-relativistic effects, but are insensitive to gravitational time dilation.

To this end, we establish a relation between special-relativistic time dilation and kinematic asymmetry of closed atom interferometers, taking the form of recoil measurements~\cite{Bouchendira11,Lan13,Parker18,Borde93}.
For these geometries, a single atomic clock in a superposition of two different trajectories undergoes special-relativistic time dilation.
The induced distinguishability leads to a loss of visibility upon interference, such that the proposed experiment represents a realisation of the twin paradox in quantum-clock interferometry.

In general relativity the proper time along a worldline $z=z(t)$ is invariant under coordinate transformations and can be approximated as
\begin{equation}\label{eq:tau}
    \tau = \int \dd \tau \cong \int \dd t \left[1- \left(\dot{z}/c\right)^2 /2 + U/c^2\right],
\end{equation}
where $c$ denotes the speed of light.
Here, $\dot{z}= \dd z/\dd t $ is the velocity of the particle and $U(z)$ the Newtonian gravitational potential along the trajectory.
This classical quantity is connected to the phase
\begin{equation} \label{eq:phi_general}
    \varphi = - \omega_{\text{C}} \tau + S_{\text{em}}/\hbar
\end{equation}
acquired by a first-quantised matter wave assuming it is sufficiently localised such that it can be associated with this trajectory.
Here, $\omega_{\text{C}}= mc^2/\hbar$ denotes the Compton frequency of a particle of mass $m$ and
\begin{equation}\label{eq:chi_general}
    S_{\text{em}} =- \int \dd t\, V_{\text{em}}
\end{equation}
is the classical action arising from the interaction of the matter wave with electromagnetic fields described by the potential $V_{\text{em}}(z,t)$ evaluated along the trajectory.
For instance, if the electromagnetic fields generate optical gratings, this potential can transfer momentum to the matter wave, thus changing its trajectory, which in turn affects proper time. 

Light-pulse interferometers~\cite{Kasevich91} use this concept of pulsed optical gratings to manipulate matter waves.
In case of interferometers closed in phase space~\cite{Roura17} and for potentials up to second order in $z$, the phase difference $\Delta \varphi$ can be calculated from Eq.~\eqref{eq:phi_general} by integrating along the classical trajectories.

\section*{Results}
\noindent\textbf{\bfseries\sffamily Time dilation and gravito-kick action} \\[1ex]
\noindent Since the light pulses act differently on the two branches of the interferometer, we add superscripts $\pathlabel=1,2$ to the potential $V_{\text{em}}^\iii$.
Moreover, we separate $V_{\text{em}}^\iii= V_{\text{k}}^\iii+V_{\text{p}}^\iii$ into a contribution $V_{\text{k}}^\iii$ causing momentum transfer and $V_{\text{p}}^\iii$ imprinting the phase of the light pulse without affecting the motional state~\cite{Schleich12}.
Consequently, we find that the motion $z^\iii= z_{\text{g}}+z_{\text{k}}^\iii$ along one branch can also be divided into two contributions: $z_{\text{g}}$ caused by the gravitational potential and $z_{\text{k}}^\iii$ determined by the momentum transferred by the light pulses on branch $\pathlabel$.

For a linear gravitational potential, the proper-time difference between both branches takes the form 
\begin{equation}\label{eq:Delta_tau}
    \Delta \tau = \int \dd t \left[ \ddot{z}_{\text{k}}^\i z_{\text{k}}^\i- \ddot{z}_{\text{k}}^\ii z_{\text{k}}^\ii\right]\Big/\left(2c^2\right)
\end{equation}
(see Materials and Methods). 
It is explicitly independent of $z_{\text{g}}$ as well as of the particular interferometer geometry, which is a consequence of the phase of a matter wave being invariant under coordinate transformations. 
When transforming to a freely falling frame, both trajectories reduce to the kick-dependent contribution $z_{\text{k}}^\iii$ and the proper-time difference $\Delta \tau$ is thus independent of gravity~\cite{Giese19}.
Accordingly, closed light-pulse interferometers are insensitive to gravitational time dilation.
Our result implies that time dilation in such interferometer configurations constitutes a purely special-relativistic effect caused by the momentum transferred through the light pulses.

Our model of atom-light interaction assumes instantaneous momentum transfer and neglects the propagation time of the light pulses.
In fact, a potential $V_{\text{k}}$ linear in $z$, where the temporal pulse shape of the light is described by a delta function, i.e. $\ddot z_{\text{k}} \propto \delta(t-t_\ell)$, reflects exactly such a transfer.
For such a potential, we find the differential action
\begin{equation} \label{eq:Delta_S_em}
    \Delta S_{\text{em}} = 2 \hbar \omega_{\text{C}} \Delta \tau + \Delta S_{\text{gk}}+ \Delta S_{\text{p}}
\end{equation}
(see Materials and Methods), which can be interpreted~\cite{Wolf11} as the laser pulses sampling the position of the atoms $z=z_{\text{k}}+z_{\text{g}}$.
The first contribution has the form of the proper-time difference, which highlights that the action of the laser can never be separated from proper time in a phase measurement in the limit given by Eq.~\eqref{eq:phi_general}.
It arises solely from the interaction with the laser and, in the case of instantaneous acceleration $\ddot z_{\text{k}}$, these kicks read out the recoil part of the motion $z_{\text{k}}$ according to Eq.~\eqref{eq:Delta_tau}.
Similarly, the second contribution in Eq.~\eqref{eq:Delta_S_em} is the action that arises from the acceleration $\ddot z_{\text{k}}$ measuring the gravitational part $z_{g}$ of the motion and takes the form
\begin{equation}\label{eq:Delta_S_kg}
    \Delta S_{\text{gk}} = m \int \dd t \,\Delta \ddot{z}_{\text{k}} z_{\text{g}},
\end{equation}
where we define the difference $\Delta \ddot{z}_{\text{k}}=\ddot{z}_{\text{k}}^\i-\ddot{z}_{\text{k}}^\ii$ between branch-dependent accelerations.
Even though this contribution is caused by the interaction with the light, the position of the atom still depends on gravity and it is caused by the combination of both the momentum transfers and gravity.
Hence, we refer to it as gravito-kick action.

Finally, the lasers imprint the laser phase action
\begin{equation}\label{eq:S_p}
\Delta S_{\text{p}} =- \int \dd t \, \Delta V_{\text{p}},
\end{equation}
with $\Delta V_{\text{p}}= V_{\text{p}}^\i- V_{\text{p}}^\ii$.

So far we have not specified the interaction with the light but merely assumed that the potential $V_{\text{k}}$ is linear in $z$.
In the context of our discussion, beam splitters and mirrors are generated through optical gratings made from two counter-propagating light beams that diffract the atoms~\cite{Kasevich91}.
In a series of light pulses, the periodicity of the $\ell$th grating is parametrised by an effective wave vector $k_\ell$.
Depending on the branch and the momentum of the incoming atom, the latter receives a recoil $\pm \hbar k_\ell$ in agreement with momentum and energy conservation.
At the same time, the phase difference of the light beams is imprinted to the diffracted atoms.

To describe this process, we employ the branch-dependent potential $V_{\text{k}}^\iii= - \sum_\ell \hbar k_\ell^\iii z^\iii \delta (t-t_\ell)$ for the momentum transfer $\hbar k_\ell^\iii$ of the $\ell$th laser pulse at time $t_\ell$ and the potential $V_{\text{p}}^\iii= - \sum_\ell \hbar \phi^\iii_\ell \delta (t-t_\ell)$ to describe the phase $\phi^\iii_\ell$ imprinted by the light pulses~\cite{Schleich12}.
Since the phases imprinted by the lasers can be evaluated trivially and are independent of $z$, we exclude the discussion of $V_{\text{p}}^\iii$ from the study of different interferometer geometries and set it to zero in the following.

\begin{figure*}
\includegraphics[width=\textwidth]{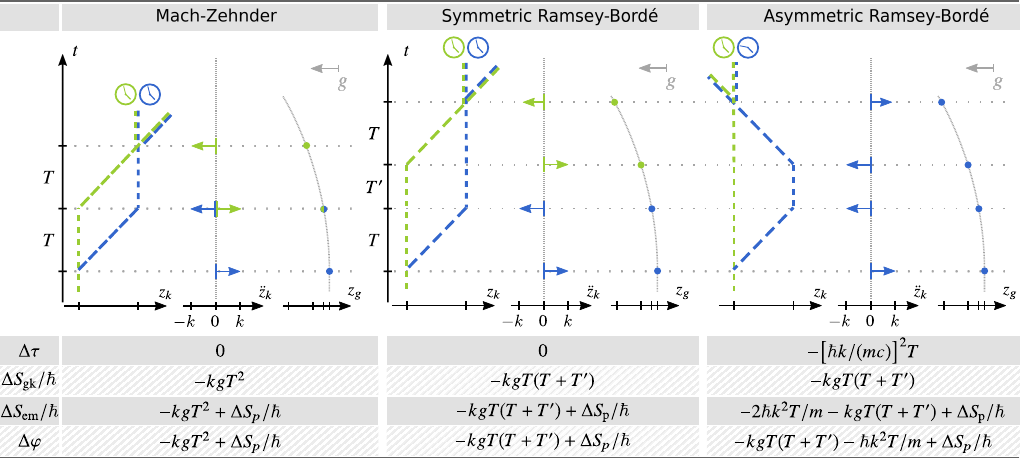}
    \caption{\textbf{\sffamily\firamedium Time dilation in different interferometer geometries.}
     Spacetime diagrams for the light-pulse and gravitationally induced trajectories $z_{\text{k}}$ and  $z_{\text{g}}$, as well as the accelerations $\ddot{z}_k$ caused by the light pulses, together with the proper-time difference $\Delta \tau$, the gravito-kick action $\Delta S_{\text{gk}}$, the electromagnetic contribution $\Delta S_{\text{em}}/\hbar$, and the total phase difference $\Delta \varphi$ of an MZI (left), a symmetric RBI (centre) and an asymmetric RBI (right).
    The first two geometries display a symmetric momentum transfer between the two branches, leading to vanishing proper-time differences. However, the asymmetric RBI features a proper-time difference that has the form of a recoil term. The spacetime diagrams also illustrate the connection to the twin paradox by displaying ticking rates (the dashes) of the two twins travelling along the two branches. Both quantum twins in the MZI and symmetric RBI experience the same time dilation, whereas in the asymmetric RBI one twin stays at rest and the other one leaves and returns so that their proper times are different. The arrows in the plot of $\ddot{z}_k$ denote the amplitude of the delta functions that scale with $\pm \hbar k/m$. Due to the instantaneous nature of $\ddot{z}_{\text{k}}$, the integration over time in Eqs.~\eqref{eq:Delta_tau} and~\eqref{eq:Delta_S_kg} reduces to a sampling of the positions $z_{\text{k}}$ and $z_{\text{g}}$ at the time of the pulses such that the respective phase contributions can be inferred directly from the figure. 
    }
    \label{fig:single_loops}
\end{figure*}

\noindent\textbf{\bfseries\sffamily Atom-interferometric twin paradox}\\[1ex]
\noindent The Kasevich-Chu-type~\cite{Kasevich91} Mach-Zehnder interferometer (MZI) has been at the center of a vivid discussion about gravitational redshift in atom interferometers~\cite{Mueller10,Wolf11,Schleich12}.
Indeed, it has been demonstrated that its sensitivity to the gravitational acceleration $g$ stems entirely from the interaction with the light, i.e. $\Delta S_{\text{gk}}$, while the proper time vanishes~\cite{Wolf11,Schleich12}.
It is hence insensitive to gravitational time dilation -- which, a priori, is not necessarily true for arbitrary interferometer geometries.

Such an MZI consists of a sequence of pulses coherently creating, redirecting and finally recombining the two branches. 
The three pulses are separated by equal time intervals of duration $T$.
We show the spacetime diagram of the two branches $z_{\text{k}}^\iii$, the light-pulse induced acceleration $ \ddot{z}_k^\iii$ as a sequence in time and the gravitationally induced trajectory $z_{\text{g}}$ in Fig.~2 on the left.

The contributions $z_{\text{k}}^{(\alpha)}$ are branch-dependent while $z_{\text{g}}$ is common for both arms of the interferometer.
From these quantities and with the help of Eqs.~\eqref{eq:Delta_tau} and~\eqref{eq:Delta_S_kg}, we obtain the phase contributions shown at the bottom of the Fig.~2 (see Materials and Methods).
The phase takes the familiar form $\Delta \varphi = -kgT^2$ and has no proper-time contribution, but is solely determined by the gravito-kick action originating in the interaction with the light pulses~\cite{Wolf11,Schleich12}.

The vanishing proper-time difference can be explained by the light-pulse induced acceleration $ \ddot{z}_{\text{k}}^\iii$ that acts symmetrically on both branches.
We draw on the classical twin paradox to illustrate the effect:
At some time, one twin starts to move away from his brother and undergoes special-relativistic time dilation, as shown by hypothetical ticking rates in the spacetime diagram (the dashing periods in Fig.~2).
After a time $T$, he stops and his brother starts moving towards him.
Because his velocity corresponds to the one that caused the separation, he undergoes exactly the same time dilation his brother experienced previously.
Hence, when both twins meet after another time interval $T$, their clocks are synchronised and no proper-time difference arises.
In an MZI, we find the quantum analogue of this configuration, where a single atom moves in a superposition two different worldlines like the quantum twin of Fig.~1B. However, only due to the symmetry of the light-pulse induced acceleration, no proper-time difference is accumulated between the branches of the interferometer.

A similar observation is made for the symmetric Ramsey-Bord\'e interferometer (RBI), where the atom separates for a time $T$, stops on one branch for a time $T'$, before the other branch is redirected.
We show the spacetime diagrams and the light-pulse induced acceleration $\ddot z_{\text{k}}$ in the center of Fig.~2 with the phase contributions below.
The two light pulses in the middle of the symmetric RBI are also beam-splitting pulses that introduce a symmetric loss of atoms.
As for the MZI, the proper-time difference between both branches vanishes and the phase is determined solely by the laser contribution and the gravito-kick phase as shown by the ticking rates in the spacetime diagram. 
The only difference with respect to the MZI is that the two branches travel in parallel for a time $T'$ during which proper time elapses identically for both of them.

The situation changes significantly when we consider an asymmetric RBI, where one branch is completely unaffected by the two central pulses as shown on the right of Fig.~2.
In fact, the twin that moved away from its initial position experiences a second time dilation on its way back so that there is a proper-time difference when both twins meet at the final pulse.
It is therefore the kinematic asymmetry that causes a non-vanishing proper-time difference, as indicated in the figure by the ticking rates. 
In fact, the proper-time difference
\begin{equation}\label{eq:phi_RBI}
    \Delta \tau_{\text{aRBI}} =- (\hbar k / m c)^2 T
\end{equation}
is proportional to a kinetic term~\cite{Lan13} that depends on the momentum transfer $\hbar k$, as already implied by Eq.~\eqref{eq:Delta_tau}.
With the light-pulse induced acceleration $ \ddot{z}^\iii_{\text{k}}$ as well as the gravitationally induced trajectory $z_{\text{g}}$ also shown on the right of the figure, we find the same contribution for $\Delta S_{\text{gk}}$ given in Fig.~2 as for the symmetric RBI.
The other contribution of $\Delta S_{\text{em}}/\hbar$ has the form $2\omega_{\text{C}} \Delta \tau_{\text{aRBI}}$ and all of them together contribute to the phase difference $\Delta \varphi$.\\

\noindent\textbf{\bfseries\sffamily Clocks in spatial superposition}\\[1ex]
\noindent While the twin paradox is helpful in gaining intuitive understanding and insight into the phase contributions, the dashing length of the worldlines in Fig.~2 only indicates the ticking rate of a \textit{hypo\-thetical} co-moving clock. 
In fact, the atoms are in a stationary internal state during propagation, whereas the concept of a clock requires a periodic evolution between two states.
As a consequence, the atom interferometer can be sensitive to special-relativistic time dilation but lacks the notion of a clock.
In a debate~\cite{Mueller10} about whether the latter is accounted for by the Compton frequency $\omega_{\text{C}}$ as the pre-factor to the proper-time difference in Eq.~\eqref{eq:phi_general}, an additional superposition of internal states~\cite{Sinha11,Zych11} was proposed.
This idea leads to an experiment where a single clock is in superposition of different branches, measuring the elapsed proper time along each branch.
In contrast to these discussions that raised questions about the role of gravitational time dilation in quantum-clock interferometry and where no specific model for the coherent manipulation of the atoms was explored~\cite{Zych11}, we have demonstrated in this article that light-pulse atom interferometers are only susceptible to special-relativistic time dilation.
In a different context, a spatial superposition of a clock has been experimentally realised, however, through an MZI geometry that is insensitive to time-dilation effects~\cite{Rosi17}.
The implementation of a twin-paradox-type experiment with an electron in superposition of different states of a Penning trap has been proposed, where the role of internal states is played by the spin~\cite{Bushev16}.
Furthermore, quantum teleportation and entanglement between two two-level systems moving in a twin-paradox geometry was considered in the framework of Unruh-DeWitt detectors~\cite{Lin15}.

To illustrate the effect of different internal states, we introduce an effective model for an atomic clock which moves along branch $\pathlabel=1,2$ in an interferometer.
In this framework~\cite{Zych17,Hu17} the Hamiltonian
\begin{equation}
    \hat{H}_j^{\pathindex} = m_j c^2+ \frac{\hat{p}^2}{2m_j} + m_j g\hat{z}+V^{\pathindex}_{\text{em}}(\hat{z},t) \quad \text{with} \quad j\in\{\statelabelExcited,\statelabelGround\}
\end{equation}
describes a single internal state of energy $E_j = m_j c^2$ with an effective potential $V^{\pathindex}_{\text{em}}$ which models the momentum transfer (see Materials and Methods).
Mass-energy equivalence in relativity implies that different internal states are associated with different energies and therefore correspond different masses $m_j$.
To connect with our previous discussion, we take the limit of instantaneous pulses neglecting the delay of the light front propagating from the laser to the atoms.
With these considerations, the Hamiltonian for a clock consisting of an excited state $\ket{\statelabelExcited}$ and a ground state $\ket{\statelabelGround}$, both forced by Bragg pulses that do not change the internal state~\cite{Giese13} on two branches $\pathlabel = 1,2$, reads
\begin{equation}\label{eq:hamiltionian}
    \hat{H}^{\pathindex} =\hat{H}_\statelabelExcited^{\pathindex} \Ket{\statelabelExcited}\Bra{\statelabelExcited}+\hat{H}_\statelabelGround^{\pathindex}\Ket{\statelabelGround}\Bra{\statelabelGround}.
\end{equation} 

Since the Hamiltonian is diagonal in the internal states, we write the time evolution along branch $\pathlabel$ as $\hat{U}^{\pathindex} = \hat{U}_a^{\pathindex} \Ket{a} \Bra{a} +\hat{U}_b^{\pathindex} \Ket{b} \Bra{b} $, where $\hat{U}_j^{\pathindex}$ is the time-evolution operator that arises from the Hamiltonian $\hat H_j^{\pathindex}$.
For an atom initially in a state $\ket{j}$ with $j=a,b$, the output state is determined by the superposition $\hat{U}_j^\i+\hat{U}_j^\ii$ and leads to an interference pattern $P_j = (1+\cos \Delta \varphi_j)/2$, where the phase difference $\Delta \varphi_j$  depends on the internal state.

In the case of quantum-clock interferometry the initial state for the interferometer is a superposition of both internal states $(\Ket{\statelabelExcited}+\Ket{\statelabelGround})/\sqrt{2}$, which form a clock that moves along both branches in superposition.
The outlined formalism shows that such a superposition leads to the sum of two interference patterns, that is $P= (P_\statelabelExcited+P_\statelabelGround)/2$.
The sum of the probabilities $P_{\statelabelExcited/\statelabelGround}$ with slightly different phases, which corresponds to the concurrent operation of two independent interferometers for the individual states, leads to a beating of the total signal and an apparent modulation of the visibility. 
Expressing the masses of the individual states by their mass difference $\Delta m$, i.e. $m_{\statelabelExcited/\statelabelGround}= m \pm \Delta m/2$, and identifying the energy difference $\Delta E =\Delta m c^2  = \hbar \Omega$ leads to the interference pattern
\begin{equation}\label{eq:P}
    P  =  \frac{1}{2}\left[1\! + \cos\!\left( \eta \frac{\Omega \Delta \tau}{2 }\right) \, \cos\!\left(\eta\omega_{\text{C}}\Delta \tau + \frac{\Delta S_{\text{gk}}+\Delta S_{\text{p}}}{\hbar} \right) \right],
\end{equation}
where the scaling factor $\eta = 1/[1- \Delta m^2/(2m)^2]$ depends on the energy difference of the two states.
In this form, the first cosine can be interpreted as a slow but periodic change of the effective visibility of the signal.
In fact, to first order in $\Delta m/m$ we find $\eta = 1$, so that the effective visibility $  \cos{(\Omega \Delta \tau/2)}$ corresponds to the signal of a clock measuring the proper-time difference.
In this picture, the loss of contrast can be seen as a consequence of distinguishability~\cite{Zych11}:
Since a superposition of internal states travels along each branch, the system can be viewed as a
clock with frequency $\Omega$ travelling in a spatial superposition.
On each branch, the clock measures proper time and by that contains which-way information, leading to a loss of visibility as a direct consequence of complementarity.

\begin{figure}
\centering
\includegraphics[width=\columnwidth]{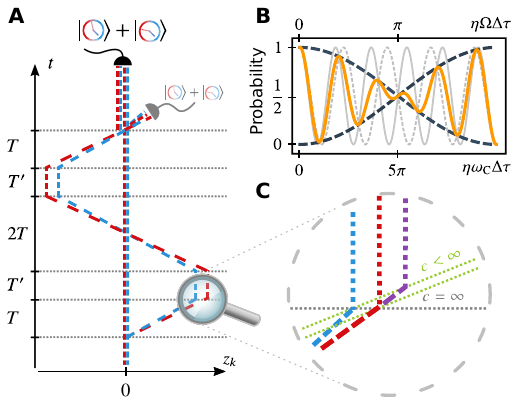}
\caption{{\sffamily\bfseries{\firamedium Interference of quantum clocks.}}
(\textbf{{\sffamily A}})~Spacetime diagram of a double-loop RBI in superposition of two different internal states (red and blue) and detection at the zero-momentum output port. We indicate the effect of different recoil velocities due to different rest masses of the internal states by slightly diverging trajectories. The different ticking rates of co-moving clocks on the trajectories are indicated by the frequency of the dashing. The dotted grey lines correspond to the light pulses used to redirect the atoms.
(\textbf{{\sffamily B}})~The output signal $P$ (solid orange) shows a visibility modulation (dashed black) which can be interpreted as the beating of the individual signals $P_{\statelabelExcited/\statelabelGround}$ of the two internal states (solid and dashed grey). To highlight the effect, we have chosen $\Delta m/m = 0.2$ in Eq.~\eqref{eq:P}. The visibility of the signal vanishes at $\eta\Omega \Delta \tau = \pi$.
(\textbf{{\sffamily C}})~Interaction of a light pulse with the excited and ground state (blue and red). Since the states follow slightly different world lines and the speed of light is finite, the light pulse will not interact simultaneously with both. Our assumption of instantaneous interaction is shown by the red and blue line. In the case of finite pulse propagation speed, indicated by the slightly titled dotted green lines, the interaction is not simultaneous and the red line for the ground state becomes the outermost purple line. 
}
\label{fig:beat}
\end{figure}

We illustrate this effect in Fig.~3A using an asymmetric double-loop RBI, where the gravito-kick action vanishes, i.e. $\Delta S_{\text{gk}}=0$, because it is insensitive to linear accelerations like other symmetric double-loop geometries that are routinely used to measure rotations and gravity gradients~\cite{Marzlin96}.
The measured phase takes the form $\Delta \varphi = 2\omega_C\tau_{\text{aRBI}} = -2\hbar k^2 T/m$.
Even though this expression is proportional to a term that has the form of proper time, it also comprises contributions from the interaction with the laser pulses, see the first term of Eq.~\eqref{eq:Delta_S_em}.
The two internal states, denoted by the blue and red ticking rates, travel along \textit{both} branches such that each twin carries its own clock, leading to a distinguishability when they meet.
This distinguishability depends on the frequency $\Omega$ of the clock and implies a loss of visibility, as shown in Fig.~3B.
However, since each internal state experiences a slightly different recoil velocity $\hbar k / m_{\statelabelExcited/\statelabelGround}$, it can be associated with a slightly different trajectory, displayed in red and blue in Fig.~3A.
The interpretation as a clock travelling along one particular branch is therefore only valid to lowest order in $\Delta m /m$.

In another interpretation, the quantum twin experiment is performed for each state independently.
The trajectories are different for each state and the proper-time difference as well as the Compton frequency are mass-dependent, so that the interferometer phase depends explicitly on the mass.
The loss of visibility can therefore be explained by the beating of the two different interference signals, which is caused by the mass difference $\Delta m = \hbar \Omega/c^2$.
In the spacetime diagram of Fig.~3A the finite speed of light pulses causing the momentum transfers is not taken into account. However, to illustrate the neglected effects induced by the propagation time, Fig.~3C magnifies such an interaction and showcases the assumption we made in our calculation: both internal states interact simultaneously and instantaneously with the light pulse, even though they might be spatially separated.
For feasible recoil velocities, as well as interferometer and pulse durations this approximation is reasonable as detailed in Materials and Methods on the light-matter interaction.

\section*{Discussion}
\noindent A realisation of quantum-clock interferometry in a twin experiment requires atomic species that feature a large internal energy splitting, suggesting typical clock atoms like strontium (Sr) with optical frequencies $\Omega$ in the order of hundreds of THz.
The proper-time difference is a property of the interferometer geometry and is enhanced for large splitting times $T$ and effective momentum transfers $\hbar k$.
Besides large-momentum-transfer techniques \cite{Parker18,Kovachy15}, this calls for atomic fountains in the order of meters~\cite{Dickerson13,Overstreet18, Hartwig15} and more or the operation in microgravity~\cite{Barrett16,Becker18}. 

To observe a full drop in visibility and its revival, the accumulated time dilation in the experiment needs to be in the order of femtoseconds. 
In the example of Sr, this can be achieved for $T=\SI{325}{ms}$ and $k=1200 \keff$, where $\keff = 1.5 \times 10^{7}\, \text{m}^{-1}$ is the effective wave number of the magic two-photon Bragg transition.
At the magic wavelength~\cite{Katori03} of $\SI{813}{nm}$, the differential ac-Stark shift of the two clock states vanishes to first order, such that the beam splitters act equally on the two internal states and hence leave the clock unaffected.
Increasing $T$ and by that $\Delta \tau$, one should observe a quadratic loss of visibility as a signature of which-path information, assuming this loss can be distinguished from other deleterious effects. 
Indeed, times up to $T = \SI{350}{ms}$ and $k=580\keff$ induce a visibility reduction of 10\%.

Although they do not use the same species, large atomic fountains~\cite{Overstreet18} already realise long free evolution times and large momentum transfer with hundreds of recoil momenta has been demonstrated~\cite{Kovachy15,Parker18}.
Techniques to compensate the impact of gravity gradients~\cite{Roura17} and rotations~\cite{Hogan08} have already proven successful~\cite{Lan12,Dickerson13,Overstreet18}.
The main challenge in implementing quantum-clock interferometers as described above lies in the concurrent manipulation of the two clock states~\cite{Rosi17}, requiring a transfer of concepts and technologies well established for alkaline atoms to alkaline earth species.
Besides magic Bragg diffraction, other mechanisms like simultaneous single-photon transitions between the clock states~\cite{Roura18} are also conceivable and relax the requirements on laser power.
In view of possible applications to gravitational wave detection~\cite{Graham16}, atom interferometry based on single-photon transitions is already becoming a major line of research.
To this end, first steps towards quantum-clock interferometry have been demonstrated by driving clock transitions of Sr to generate MZI geometries ~\cite{Hu17}.

Because the effect can be interpreted as a beating of the signal of two atomic species (defined through their internal state), one can also determine the phase for each state independently and infer their difference in the data analysis.
A differential phase of \SI{1}{mrad} assuming $T=\SI{60}{ms}$ and $k=70\keff$ may already be resolved in a table-top setup in a few hundred shots with $10^6$ atoms, supposing shot-noise limited measurements of the two internal states.
Equation~\eqref{eq:Delta_tau} shows that proper-time differences in our setting arises only from special-relativistic effects caused by the momentum transfer.
Such an experiment is equivalent to the comparison of two recoil measurements~\cite{Bouchendira11,Parker18} performed independently but simultaneously to suppress common-mode noise.
Beyond recoil spectroscopy, state resolving measurements can be of particular interest for a doubly differential measurement scheme that, in contrast to the setup discussed above, does not rely on an initial superposition of two internal states.
Instead, the superposition of internal states is generated during the interferometer~\cite{Roura18}, such that these setups can be used to measure the time dilation caused by a gravitational redshift.
In contrast, our discussion highlights the relevance of special-relativistic time dilation for the interference of quantum clocks in conventional interferometers without internal transitions.

In summary, we have shown that for an interferometer that does not change the  internal state during the sequence, the measured proper-time difference is in lowest order independent of gravity and is non-vanishing only in recoil measurements, connecting matter-wave interferometry to the special-relativistic twin paradox.
As a consequence of this independence, such light-pulse atom interferometers are insensitive to gravitational time dilation.

The light pulses creating the interferometer cause a contribution to its phase that is of the same form as the special-relativistic proper-time difference and depends on the position of the branches in a freely falling frame, which can be associated with the worldline of a quantum twin.
Since these trajectories and by that proper time depends on the recoil velocity that is slightly different for different internal states, an initial superposition causes a beating of two interference patterns.
In such a quantum version of the twin-paradox, a clock is in a spatial supersposition of different worldlines, leading to a genuine implementation of quantum-clock interferometry but based on special-relativistic time dilation only.

\section*{Materials and Methods}
\noindent\textbf{\bfseries\sffamily Recoil terms and proper time}\\[1ex]
\label{supp:recoilandpropertime}
\noindent In this section, we show that for light pulses acting instantaneously on both branches and gravitational potentials up to linear order, the proper time consists only of recoil terms.
We provide the explicit expressions for the proper-time difference and find a compact form for the action of the electromagnetic potential describing pulsed optical gratings that contributes to the phase of the atom interferometer.

As already implied by the decomposition from Eq.~\eqref{eq:chi_general}, the interaction of an atom with a light pulse transfers momentum and imprints a phase on the atom~\cite{Schleich12}.
Since the latter contribution does not modify the motion of the atom, we find $\partial V_{\text{p}}/\partial z =0$.
Consequently, the classical equations of motion can be written as
$
    m \ddot{z}= - \partial (m U)/\partial z-\partial V_{\text{k}}/\partial z= m \ddot{z}_{\text{g}}+m \ddot{z}_{\text{k}}.
$
The integration of these equations leads to the trajectory $z=z(t)$ that can be decomposed into two contributions associated with these accelerations, i.e. $z= z_{\text{g}}+z_{\text{k}}$, where we collect the initial conditions in $z_{\text{g}}$.

Proper time takes in lowest order expansion in $c^{-2}$, i.e. for weak fields and low velocities, according to Eq.~\eqref{eq:tau} for a linear gravitational potential the form
\begin{equation}
    c^2 \tau = \int \dd t \left(c^2- \dot{z}^2 /2 - \ddot{z}_{\text{g}} z\right).
\end{equation}
We simplify this expression by integrating the kinetic term $\dot{z}^2 /2$ by parts and make the substitution $z= z_{\text{g}}+z_{\text{k}}$ in the remaining integral, so that we find
\begin{equation}
    c^2 \tau =  \left.-\frac{\dot{z}z}{2}\right| +  \int \dd t \left(c^2 +\frac{\ddot{z}_{\text{k}}z_{\text{k}}}{2} -\frac{\ddot{z}_{\text{g}}z_{\text{g}}}{2} + \frac{\ddot{z}_{\text{k}}z_{\text{g}}-\ddot{z}_{\text{g}}z_{\text{k}}}{2}\right) 
\end{equation}
for the proper time. Partial integration of the last term in the integral leads to the compact form
\begin{equation}\label{eq:integrated_tau}
    c^2 \tau =   \left.\frac{\dot{z}_{\text{k}}z_{\text{g}} -\dot{z}_{\text{g}}z_{\text{k}} -\dot{z}z}{2}\right| +  \int \dd t \left(c^2 -\frac{\ddot{z}_{\text{g}}z_{\text{g}}}{2}+  \frac{\ddot{z}_{\text{k}}z_{\text{k}}}{2}\right)  
\end{equation}
that explicitly depends on the initial and final positions and velocities.

In a light-pulse atom interferometer, light pulses act independently through the potentials $V_{\text{k}}^\iii$ on the two branches $\alpha=1,2$ and give rise to the light-pulse induced trajectories $z_{\text{k}}^\iii$.
In turn, these branch-dependent potentials lead to a proper-time difference $\Delta \tau = \tau^\i-\tau^\ii$ between the upper and lower branch of the interferometer and cause a phase contribution to the interference pattern.
In an interferometer closed in phase space, the initial and final positions as well as velocities are the same for both branches and thus the first term in Eq.~\eqref{eq:integrated_tau} vanishes.
Since $z_{\text{g}}$ is branch-independent, the first two terms in the integral cancel as well and we are left with
\begin{equation}\label{eq:Delta_tau_suppl}
    \Delta \tau =  \int \dd t \left[ \ddot{z}_{\text{k}}^\i z_{\text{k}}^\i- \ddot{z}_{\text{k}}^\ii z_{\text{k}}^\ii \right]\Big/\left(2c^2\right)
\end{equation}
for the lowest order of the proper-time difference of an atom interferometer in a linear gravitational potential.
In fact, the proper-time difference in a closed interferometer is independent of gravity and constitutes a special-relativistic effect.
This result can also be derived for a time-dependent gravitational acceleration $g(t)$.

Since proper time is invariant under coordinate transformations, the proper-time difference of a closed atom interferometer is independent of the gravitational acceleration by considering the common freely falling frame.
In this frame, the trajectories are straight lines and correspond to $z_{\text{k}}^\iii$, as implied by Fig.~2, so that the proper-time difference is of special relativistic origin.
Hence, Eq.~\eqref{eq:Delta_tau_suppl} can be also interpreted as a direct consequence of transforming to a freely falling frame in a homogeneous gravitational field. 

The laser contribution to the phase can be calculated from Eq.~\eqref{eq:chi_general} and we write the classical action in the form of
\begin{equation}
    S_{\text{em}} =- \int \dd t \left(V_{\text{k}}+V_{\text{p}}\right) =  m \int \dd t \ddot{z}_{\text{k}} z -\int \dd t V_{\text{p}},
\end{equation}
where we assumed that $V_{\text{k}}= - m \ddot{z}_{\text{k}} z$ is linear in $z$. 
When we again use the decomposition of the position $z= z_{\text{g}}+z_{\text{k}}$ into a part induced by gravity and a part induced by the light pulses, we find
\begin{equation}
    S_{\text{em}} =  m\int \dd t  \ddot{z}_{\text{k}} z_{\text{k}}+ m\int \dd t  \ddot{z}_{\text{k}} z_{\text{g}}-\int \dd t V_{\text{p}}
\end{equation}
for the action.
Since $z_{\text{g}} $ is branch-independent in contrast to $z_{\text{k}}$, the difference
\begin{equation}
    \Delta S_{\text{em}} = m  \int \dd t   \left[ \ddot{z}_{\text{k}}^\i z_{\text{k}}^\i- \ddot{z}_{\text{k}}^\ii z_{\text{k}}^\ii \right]+ m\int \dd t \, \Delta \ddot{z}_{\text{k}} z_{\text{g}}-\int \dd t\, \Delta V_{\text{p}}
\end{equation}
between upper and lower branch depends on $\Delta \ddot{z}_{\text{k}}= \ddot{z}_{\text{k}}^\i-\ddot{z}_{\text{k}}^\ii$ and  $\Delta V_{\text{p}}=V_{\text{p}}^\i-V_{\text{p}}^\ii$.
With the expression for the proper time from Eq.~\eqref{eq:Delta_tau}, the gravito-kick action from Eq.~\eqref{eq:Delta_S_kg} and the laser phase action from Eq.~\eqref{eq:S_p} we arrive at the form of Eq.~\eqref{eq:Delta_S_em} for the action of the interaction with the electromagnetic field such as pulsed optical gratings.

For the specific form of the phase contributions and proper time, we first calculate the trajectory that arises from $\ddot{z}_{\text{g}}= - g$ and find by simple integration
\begin{equation}\label{eq:z_g}
    z_{\text{g}}(t)= z (0) + \dot{z}(0) \,t - g t^2/2 ,
\end{equation}
which is branch-independent.
For the specific form of the phase contributions and proper time, we calculate the trajectories that arise from the atom-light interaction.
To this end, we assume the potential $V_{\text{k}}^\iii= - \sum_\ell \hbar k_\ell^\iii z^\iii \delta (t-t_\ell)$ that causes the momentum transfer $m\ddot{z}_{\text{k}}^\iii=\sum_\ell \hbar k_\ell^\iii  \delta (t-t_\ell)$.
Integrating the acceleration leads to the two branches of the interferometer given by the two trajectories
\begin{equation}\label{eq:z_k}
    z_{\text{k}}^\iii (t) = \sum\limits_{\ell=1}^{n} (t-t_\ell) \hbar k_\ell^\iii /m 
\end{equation}
for $t_{n+1} > t> t_{n}$.

Using the expression for $\ddot{z}_{\text{k}}^\iii$, the gravito-kick phase from Eq.~\eqref{eq:Delta_S_kg} takes the explicit form
\begin{equation}
    \Delta S_{\text{gk}}/\hbar = \sum\limits_\ell \Delta k_\ell z_{\text{g}}(t_\ell),
\end{equation}
where we evaluate the gravitationally induced trajectory from Eq.~\eqref{eq:z_g} at the times of the pulses.
Note that we defined the differential momentum transfer $\Delta k_\ell = k_\ell^\i-k_\ell^\ii$ of the $\ell$th laser pulse.

With the branch-dependent trajectory from Eq.~\eqref{eq:z_k} and  $\ddot{z}_{\text{k}}^\iii$, we perform the integration in Eq.~\eqref{eq:Delta_tau_suppl} to arrive at the expression
\begin{equation}
    \omega_{\text{C}} \Delta \tau = \frac{\hbar}{2m} \sum\limits_{n=1}^{M} \sum\limits_{\ell=1}^n \left[k_n^\i k_\ell^\i-k_n^\ii k_\ell^\ii \right] (t_n-t_\ell)\,,
\end{equation}
where $M$ is the total number of light-matter interaction points.
This phase difference is proportional to $\hbar / m$ and includes a combination of the transferred momenta and separation between the laser pulses.\\

\noindent\textbf{\mdseries\sffamily Post-Galilean bound systems in a Newtonian gravitational field}\\[1ex]
We consider a static spacetime with a line element of the form 
\begin{equation}\label{eq.app.statmetric}
    \text{d}s^2 = g_{\mu\nu}(x)\text{d}x^{\mu}\text{d}x^{\nu} = -N(\vec{x}) (c\mathrm{d}t)^2 + B_{ij}(\vec{x})\text{d}x^{i}\text{d}x^{j}
\end{equation}
where $g_{\mu\nu}$ is the metric with Greek indices running from $0$ to $3$, $N$ is the lapse function and $B_{ij}$ is the three-metric with Latin indices from $1$ to $3$.
In the limit of Newtonian gravity the lapse function becomes $N(\vec{x})=1+2U(\vec{x})/c^2$ with $U$ being the Newtonian gravitational potential while $B_{ij}=\big[1-2U(\vec{x})/c^2\big]\delta_{ij}$, where $\delta_{ij}$ is the Kronecker symbol. The post-Newtonian correction to the spatial part of the metric decomposition only needs to be considered for the electromagnetic field but not for the atoms inside a light-pulse atom interferometer. 

To model atomic multi-level systems inside such a background metric including effects that arise from special-relativistic and general-relativistic corrections due to the lapse function of the metric to the bound state energies, one can resort to a quantum field theoretical treatment~\cite{Anastopoulos18} and perform the appropriate limit to a first-quantised theory afterwards.
In this approach one first performs the second quantisation of the respective interacting field theory in the classical background metric provided by Eq.~\eqref{eq.app.statmetric} and derives the bound state energies as well as possible spin states of e.g. hydrogen-like systems.
For each pair of energy and corresponding internal state, one can take the limit of a first-quantised theory and Newtonian gravity.
Expanding the Newtonian gravitational potential up to second order leads to a Hamiltonian
\begin{equation}\label{eq.app.H_0j}
    \hat{H}_{0,j}=  m_j c^2 + \frac{\hat{\vec{p}}^2}{2m_j} + m_j\left( \vec{g}^\intercal \hat{\vec{x}} + \frac{1}{2} \hat{\vec{x}}^\intercal\Gamma \hat{\vec{x}}\right) .
\end{equation}
Here, $E_j= m_jc^2$ is the energy corresponding to the energy eigenstate $\ket{j}$, $\hat{\vec{p}}$ and $\hat{\vec{x}}$ are the momentum and position operator, respectively, $\vec{g}$ is the (local) gravitational acceleration vector and $\Gamma$ is the (local) gravity gradient tensor.
This Hamiltionian includes special-relativistic and possibly post-Newtonian contributions to its internal energies as indicated by the different masses $m_j$ of the individual internal states, which is a direct manifestation of the mass-energy equivalence.
In principle, terms proportional to  $\hat{\vec{p}}^4$ and $\hat{\vec{p}}^\intercal(\vec{g}^\intercal\hat{\vec{x}}+\frac{1}{2}\hat{\vec{x}}^\intercal \Gamma\hat{\vec{x}})\,\hat{\vec{p}}$ appear as a correction to the centre-of-mass Hamiltonian.
However, since these terms are state independent to order $1/c^2$, they leave the beating in Eq.~\eqref{eq:P} unaffected and will therefore be disregarded.
The third addend in Eq.~\eqref{eq.app.H_0j} is the Newtonian gravitational potential energy which we denote by $V_{\text{g}}(\hat{\vec{x}})$. 
The fact that each state couples separately to the (expanded) gravitational potential with its respective mass $m_j$ directly highlights the weak equivalence principle.
The full Hamiltonian of an atomic system with multiple internal states labelled by the index $j$ is thus $\hat{H}_0 = \sum_{j} \hat{H}_{0,j} \ket{j}\bra{j}$. This Hamiltonian is diagonal with respect to different internal states and gravity induces no cross-coupling between the states of freely-moving atoms.\\

\noindent\textbf{\sffamily Light-matter interaction and total Hamiltonian}\\[1ex]
In typical light-pulse atom interferometers the light-matter interaction is only switched on during the beam-splitting pulses.
Hence, the propagation of the atoms through an interferometer can be partitioned into periods of free propagation and periods where the lasers are acting on the atoms.
In particular, the light-matter interaction including post-Newtonian corrections to the atoms' bound state energies~\cite{Marzlin95,Sonnleitner2018} in the low-velocity, dipole approximation limit reduces to
\begin{equation}
    \hat{\mathcal{V}}_{\text{em}}(\hat{\vec{x}},t) = -\hat{\vec{\wp}} ~\vec{E}(\hat{\vec{x}},t) + \hat{\mathcal{V}}_{\text{R}}(\hat{\vec{x}},t;\hat{\wp})
\end{equation}
where $\hat{\vec{\wp}}$ is the electric dipole moment operator, $\vec{E}$ is the external electric field and $\hat{\mathcal{V}}_{\text{R}}$ is the R\"ontgen contribution to the interaction Hamiltionian.
The information about special-relativistic corrections to the energies is included in the definition of the dipole moments.
Moreover, since the light-matter coupling is of the usual form we can apply the standard framework of quantum optics to derive effective models for the interaction.

However, before proceeding we simplify our model by only considering unidirectional motion in the $z$-direction and an acceleration $g$ anti-parallel to it, so that the Hamiltionian for the full interferometer becomes
\begin{align}
\begin{split}\label{eq.app.H_0} 
        \hat{H}(\hat{z},t) = \sum_j \left[m_j c^2 + \frac{\hat{p}^2}{2m_j} + m_j\left(g \hat{z} + \frac{\Gamma}{2} \hat{z}^2\right)\right]\ket{j}\bra{j} \\ + \sum_{i\neq j} \mathcal{V}_{\text{em},ij}(\hat{z},t)\ket{i}\bra{j},
\end{split}
\end{align}
where $\mathcal{V}_{\text{em},ij}$ are the time-dependent matrix elements of the light-matter coupling which include the switch-on/off of the lasers. 
Based on this model we can derive an effective potential description for e.g. two-photon Raman or Bragg transitions inside an interferometer.
In particular, for magic Bragg diffraction we consider pairs of one relevant state $\ket{j}$ and one ancilla state out of the atomic state manifold, each interacting with two light fields.
After applying the rotating wave approximation, adiabatic elimination, two-photon resonance conditions~\cite{Giese13} and taking the limit of instantaneous pulses, we can replace the electromagnetic interaction by the effective potential
\begin{equation}\label{eq:lightmatter-pulses-simultaneous}
    \hat{V}_{\text{em}}= -\hbar\sum_{j} \Big(k_{\ell}\hat{z}+\phi_{\ell}\Big)\delta(t-t_\ell)\ket{j} \bra{j}.
\end{equation}
Here we evaluate the effective momentum transfer $\hbar k_{\ell}$ as well as the phase $\phi_{\ell}$ of the electromagnetic field at time $t_\ell$ of the pulse.
Although we have written the state dependence explicitly in Eq.~\eqref{eq:lightmatter-pulses-simultaneous}, the effective interaction $\hat{V}_{\text{em}}$ becomes state-independent since the sum over the relevant states $\ket{j}$ corresponds to unity in case of magic Bragg diffraction.

During a typical light-pulse atom interferometer sequence one usually has multiple wave-packet components centred on different trajectories, each of which constitute an individual branch of the interferometer.
In this case, the previously defined interaction can be applied \cite{Roura17} on each branch individually. 
Hence, the effective interaction Hamiltionan in the case of instantaneous Bragg pulses becomes
\begin{equation}\label{eq:lightmatter-pulses-simultaneous-branches}
    \hat{V}_{\text{em}}^{\pathindex}=-\hbar \sum_{\ell} \Big(k_{\ell}^{\pathindex} \hat{z}+\phi_{\ell}^{\pathindex}\Big)\delta(t-t_\ell),
\end{equation}
where the superscript $\pathlabel$ labels the individual branch.
A perturbative treatment shows that the approximation of instantaneous pulses is appropriate if the interaction time of the laser pulses with the atoms is sufficiently short compared to the duration of the interferometer~\cite{Bertoldi19}.
Furthermore, the propagation delay of the lightfront between wave-packet components introduces a further phase contribution. However, this phase is suppressed in the differential measurement by an additional factor of $\hbar k/(m_{a/b}c)$ compared to the phases of interest, it is thus of order $1/c^3$ and can be neglected.
\\

\noindent\textbf{\sffamily Branch-dependent light-pulse atom interferometry}\\[1ex]
As shown above, the momentum transfer caused by light pulses can be described by an effective potential that in general depends on the classical trajectory of the particle.
In our limit this dependence reduces to a mere dependence on the two branches of an atom interferometer.
Since our description is diagonal in the different internal states and we assume that throughout the free propagation inside the interferometer the internal state of the atoms does not change, the time evolution along branch $\alpha=1,2$ takes the form
\begin{equation}
     \hat{U}^{\pathindex} = \hat{U}_\statelabelExcited^{\pathindex} \Ket{\statelabelExcited} \Bra{\statelabelExcited} +\hat{U}_\statelabelGround^{\pathindex} \Ket{\statelabelGround} \Bra{\statelabelGround},
\end{equation}
where $\hat{U}_j^{\pathindex}$ is the time-evolution operator for state $\ket{j}$ along path $\alpha$ ending in one particular exit port.
We limit our discussion to one excited state and one ground state, hence we use the labels $j=\statelabelExcited,\statelabelGround$, respectively.
If $\ket{\psi_j(0)}$ describes the initial external degree of freedom of the atoms in state $\ket{j}$ and we project onto the internal state when we perform the measurement, the postselected state in one of the output ports of the interferometer is a superposition of the two branches, i.e. $\ket{\psi_j}= \big(\hat{U}_j^{\i}+\hat{U}_j^\ii\big)\ket{\psi_j(0)}/2$, leading to the interference pattern
\begin{equation}
    P_j = \braket{\psi_j|\psi_j}= \frac{1}{2}\Big(1+ \frac{1}{2}\braket{\psi_j(0)|\hat{U}_j^{(2)\dagger} \hat{U}_j^{(1)}|\psi_j(0)}+ \text{c.c.}\Big).
\end{equation}
The calculation of the inner product can now be performed using the explicit form of the branch-dependent potentials.
For a closed geometry and potentials up to linear order, the calculation reduces to the description outlined in the main part of the article~\cite{Schleich12}.
This treatment is also exact in the presence of gravity gradients and rotations but will lead in general to open geometries, which can be closed through suitable techniques~\cite{Roura17}.
When we introduce the state-dependent Compton frequency $\omega_j = m_j c^2/\hbar$ and proper-time difference $\Delta \tau_j$ where $m$ has to be replaced by $m_j$, we find
$ P_j= \left(1 + \cos \Delta \varphi_j \right)/2$ and the phase difference
\begin{equation}\label{eq.varphi_j}
    \Delta \varphi_j = - \omega_j \Delta \tau_j + 2 \omega_j \Delta \tau_j + \Delta S_{\text{gk}} /\hbar + \Delta S_{\text{p}}/\hbar.
\end{equation}
Here, we used the fact that both $\Delta S_{\text{gk}}$ and $\Delta S_{\text{p}}$ do not depend on the internal state in accordance with the weak equivalence principle.
Moreover the phases are degenerate if the proper-time difference vanishes.

If the atoms are initially in a superposition of the two internal states, i.e. $(\ket{\statelabelExcited}+\ket{\statelabelGround})/\sqrt{2}$, the exit port probability without postselection on one internal state is $P= (P_\statelabelExcited+P_\statelabelGround)/2$, which corresponds to the sum of the two interference patterns.
After some trigonometry we find
\begin{equation}\label{eq.app.beating}
    P = \frac{1}{2}\left[1+ \cos\left(\frac{\Delta \varphi_\statelabelExcited -\Delta \varphi_\statelabelGround }{2}\right) \cos\left(\frac{\Delta \varphi_\statelabelExcited + \Delta \varphi_\statelabelGround }{2}\right) \right]
\end{equation}
so that the two interference patterns beat.
The first term, i.e. the difference of the phases of the individual states, can be interpreted as a visibility modulation of the concurrent measurement.
Because the two masses $m_{\statelabelExcited/\statelabelGround}= m \pm \Delta m/2$ are connected to the energy difference $\Delta E = \hbar \Omega = \Delta m c^2 $ between excited and ground state, the frequency $\Omega$ determines the beating.\\

\noindent\textbf{\sffamily Connection to clock Hamiltonians}\\[1ex]
We discuss in the main body of the article that in an expansion of the phase difference in orders of $\Delta m /m$, the beating effect can be interpreted as a loss of contrast due to the distinguishability of two internal clock states.
In this section we show the connection of the Hamiltonian from Eq.~\eqref{eq.app.H_0} to a clock Hamiltonian~\cite{Zych11}
\begin{equation}\label{eq.app.H_int}
    \Hat{H}_{\text{int}}= \Big(m c^2 -\frac{\hbar \Omega}{2}  \Big) \Ket{\statelabelGround}\Bra{\statelabelGround} +  \Big(m c^2 +\frac{\hbar \Omega}{2}  \Big) \Ket{\statelabelExcited}\Bra{\statelabelExcited}.
\end{equation}
In fact, expanding Eq.~\eqref{eq:hamiltionian} up to linear order of $\Delta m / m$, we find the expression
\begin{equation}
    \hat{H}^{\pathindex} = \Hat{H}_{\text{int}}+ \frac{\hat{p}^2}{2m} + m g\hat{z}+V^{(\alpha)}_{\text{em}}+  \left(-\frac{\hat{p}^2}{2m} + m g\hat{z} \right) \frac{(\Hat{H}_{\text{int}}-mc^2\mathbbm{1}_2)}{mc^2}
\end{equation}
with the help of Eq.~\eqref{eq.app.H_int}.
In this form, the coupling of the internal dynamics to the external degrees of freedom is prominent and leads to the interference signal from Eq.~\eqref{eq:P} with $\eta=1$.
Hence, the Hamiltonian describes a moving clock experiencing time dilation~\cite{Zych11}.\\[2ex]

\noindent{\bfseries \sffamily Acknowledgments:}
We thank B.~L.~Hu for bringing \cite{Lin15} to our attention. \\[1ex]

\noindent{\bfseries \sffamily Funding:}
We acknowledge financial support from DFG through CRC 1227 (DQ-mat), project B07.
The presented work is furthermore supported by CRC 1128 geo-Q, the German Space Agency (DLR) with funds provided by the Federal Ministry of Economic Affairs and Energy (BMWi) due to an enactment of the German Bundestag under Grant No. 50WM1641, 50WM1556 (QUANTUS IV), 50WM1956 (QUANTUS V) and50WM0837, as well as by ``Niedersächsisches Vorab'' through the ``Quantum- and Nano- Metrology (QUANOMET)'' initiative within the project QT3 and through ``Förderung von Wissenschaftund Technik in
Forschung und Lehre'' for the initial funding of research in the new DLR-SI Institute.
The work of IQ\textsuperscript{ST} is financially supported by the Ministry of Science, Research and Arts Baden-W\"urttemberg.
DS gratefully acknowledges funding by the Federal Ministry of Education and Research (BMBF) through the funding program Photonics Research Germany under contract number 13N14875.
M.Z. acknowledges support through ARC DECRA grant no. DE180101443 and ARC Centre EQuS CE170100009.
W.P.S. thanks Texas A{\&}M University for a Faculty Fellowship at the Hagler Institute for Advanced Study at Texas A{\&}M University and Texas A{\&}M AgriLife for the support of this work.\\[1ex]
{\bfseries \sffamily Author contributions:}
All authors contributed to scientific discussions, the execution of the study, and the interpretation of the results.
S.L. and A.F. contributed equally to this work.
S.L., A.F., C.U., F.D.P. and E.G. prepared the manuscript with input from all other authors.
E.M.R., W.P.S. and E.G. supervised the project. \\[1ex]
{\bfseries \sffamily Competing interests:}
All authors declare that they have no competing financial and nonfinancial interests. \\[1ex]
{\bfseries \sffamily Data and materials availability:}
All data generated or analyzed during this study are included in this published article or are available from the corresponding author on reasonable request.

\bibliography{twin_sci_adv}

\end{document}